\documentclass[twoside,10pt]{article}

\usepackage{multicol}
\usepackage{amsmath}
\usepackage{amsfonts}
\usepackage{cite}
\usepackage{epsf,psfig,graphicx}

\begin{document}

\title{Boundary-Layer Theory, Strong-Coupling Series,\\
and Large-Order Behavior}

\author{Carl M. Bender \\
Department of Physics, Washington University \\
St.~Louis MO 63130, USA \\
cmb@howdy.wustl.edu \\
\and
Axel Pelster \\
Institut f\"ur Theoretische Physik \\
Freie Universit\"at Berlin, Arnimallee 14, 14195 Berlin, Germany \\
pelster@physik.fu-berlin.de
\and
Florian Weissbach \\
Institut f\"ur Theoretische Physik \\
Freie Universit\"at Berlin, Arnimallee 14, 14195 Berlin, Germany \\
weisbach@physik.fu-berlin.de}

\date{Submitted to Journal of Mathematical Physics \\
11 March 2002}

\maketitle


\typeout{==> Abstract}
\begin{abstract}
\noindent
The introduction of a lattice converts a singular boundary-layer problem in
the
continuum into a regular perturbation problem. However, the continuum limit
of
the discrete problem is extremely nontrivial and is not completely
understood.
This paper examines two singular boundary-layer problems taken from
mathematical
physics, the instanton problem and the Blasius equation, and in each case
examines two strategies, Pad\'e resummation and variational perturbation
theory,
to recover the solution to the continuum problem from the solution to the
associated discrete problem. Both resummation procedures produce good and
interesting results for the two cases, but the results still deviate from
the
exact solutions. To understand the discrepancy a comprehensive large-order
behavior analysis of the strong-coupling lattice expansions for each of the
two
problems is done.
\end{abstract}

\typeout{==> Section 1}
\section{Introduction}
\label{sec1}

In this paper we report some major advances in understanding (albeit not a
complete solution to) a difficult general class of problems in
mathematical physics. We consider here the conversion of a continuum problem
into a discrete problem by the insertion of a lattice spacing parameter $a$,
the
solution of the continuum problem on the lattice, and the subsequent
extremely
subtle continuum limit $a\to0$.

Almost every continuum physics problem is singular as a function of the
parameters in the problem. As a result, only rarely does the perturbation
series take the form of a Taylor series having a nonzero radius of 
convergence. As an
elementary example, consider the algebraic polynomial equation
\begin{eqnarray}
\epsilon x^3+x-1=0.
\label{eq1}
\end{eqnarray}
This problem is singular in the limit $\epsilon\to0$. In this limit, the
degree
of the polynomial changes from three to one and thus two of the roots
abruptly
disappear. As a consequence, a perturbative solution to this problem
[expressing
the roots $x(\epsilon)$ as series in powers of $\epsilon$] yields
expressions that are more complicated than Taylor series.

A more elaborate example of a singular problem is the time-independent
Schr\"odinger equation
\begin{eqnarray}
-\frac{\hbar^2}{2M}\nabla^2\Psi({\bf x})+[V({\bf x})-E]\Psi({\bf x})=0.
\label{eq2}
\end{eqnarray}
In the classical limit $\hbar\to0$ this differential equation abruptly
becomes
an {\it algebraic} equation, and thus the general solution no longer
contains
any arbitrary constants or functions and, as a result, it can no longer
satisfy
the initial conditions. We know that for small $\hbar$ the solution is not
Taylor-like but rather is a singular exponential in WKB form:
\begin{eqnarray}
\Psi({\bf x})\sim e^{S({\bf x})/\hbar}\quad(\hbar\to0).
\label{eq3}
\end{eqnarray}

In the study of quantum field theory, it is well known that infinities
appear in
the perturbative expansion in powers of the coupling constant. There are two
kinds of infinities. The first kind, which is due to the point-like nature
of
the interaction, requires the use of renormalization. The second kind, which
is
due to singularities in the complex-coupling-constant plane, forces the
perturbation series to have a zero radius of convergence.

A quantum field theory can be regulated by introducing a lattice spacing.
The
resulting discrete theory is completely finite and can be studied
numerically by
using various kinds of numerical methods such as Monte Carlo integration.
However,
the underlying singular nature of the continuum quantum field theory
resurfaces
in the continuum limit $a\to0$. The introduction of a lattice spacing and
the
singular nature of the continuum limit was investigated in a series of
papers by
Bender {\it et al.}~\cite{Ben1,Ben2,Ben3,Ben4,Ben5,Ben6,Ben7,Ben8,Ben9}.

A quantum field theory is just one instance in which discretization
regulates
and eliminates the singular nature of the problem. It is also known that
introducing a lattice spacing converts a boundary-layer problem, which is a
singular perturbation problem, into a regular perturbation problem
\cite{Bender1,Bender2,Bender3}. A {\it boundary-layer problem} is a
differential-equation-boundary-value problem in which the highest derivative
of
the differential equation is multiplied by a small parameter $\epsilon$.
Consider as an example
\begin{eqnarray}
\epsilon y''(x)+a(x)y'(x)+b(x)y(x)=c(x),
\label{eq4}
\end{eqnarray}
where
the boundary conditions on the function $y(x)$ typically have a form such as
\begin{eqnarray}
y(0)=A,\qquad y(1)=B.
\label{eq5}
\end{eqnarray}
This boundary-value problem is singular because in the limit $\epsilon\to0$
one of the solutions abruptly disappears and the limiting solution is not
able
to satisfy the two boundary conditions in (\ref{eq5}). The usual way to
solve
the boundary-value problem (\ref{eq4}) -- (\ref{eq5}) is to decompose the
interval $0\leq x\leq1$ into two regions, an {\it outer region}, in which
the
solution varies slowly as a function of $x$, and an {\it inner region} or
{\it
boundary-layer region}, in which the solution varies rapidly as a function
of
$x$. The boundary-layer region is a narrow region whose thickness is
typically
of order $\epsilon$ or some power of $\epsilon$ \cite{BO}.

An important example of a boundary-layer problem is the instanton equation
\begin{eqnarray}
\epsilon^2 f''(x)+f(x)-f^3(x)=0,
\label{BL3}
\end{eqnarray}
with the associated boundary conditions
\begin{eqnarray}
\label{BOO}
f(0)=0,\qquad f(\infty)=1.
\end{eqnarray}
The exact solution to this instanton problem is
\begin{eqnarray}
\label{BL4}
f(x)=\tanh \frac{x}{\epsilon\sqrt{2}}.
\end{eqnarray}
Note that the solution $f(x)$ varies rapidly at the origin $x=0$ over a
region
of thickness $\epsilon$; this is the boundary-layer region. The solution
varies
slowly (it is approximately $1$) outside of this region. The outer region
consists of those $x$ not near the origin.

A novel way to solve the instanton problem is to discretize it by
introducing a
lattice. On the lattice, the differential equation becomes a difference
equation
that can easily be solved perturbatively. In the continuum limit, as the
lattice
spacing vanishes, we then obtain a strong-coupling expansion that must be
evaluated by means of a Pad\'e or a variational perturbation theory method.
To
illustrate the approach our objective will be to calculate the slope of the
instanton at $x=0$, which from (\ref{BL4}) has the value
\begin{eqnarray}
\label{BL4a}
f'(0)=\frac{1}{\epsilon\sqrt{2}}.
\end{eqnarray}

We introduce a lattice with lattice spacing $a$ so that the real axis is
discretized in steps of width $a$. The spatial coordinate reads $x_n=na$,
where
the function $f(x)$ assumes the value $f_n=f(x_n)$. On the lattice the
second
spatial derivative in (\ref{BL3}) becomes
\begin{eqnarray}
\label{BL5}
f''(x)~\rightarrow~\frac{f_{n+1}-2f_n+f_{n-1}}{a^2}.
\end{eqnarray}
Thus, from the instanton equation (\ref{BL3}) we obtain the difference
equation
\begin{eqnarray}
\label{BL6}
\frac{\epsilon^2}{a^2}(f_{n+1}-2f_n+f_{n-1})+f_n-f_n^3=0,
\end{eqnarray}
where the boundary values follow from (\ref{BOO}):
\begin{eqnarray}
\label{BL6a}
f_0=0,\qquad f_\infty=1.
\end{eqnarray}

The natural expansion parameter now is $\epsilon^2/a^2$, to which we assign
the
name $\delta$:
\begin{eqnarray}
\label{delta}
\delta\equiv\frac{\epsilon^2}{a^2}.
\end{eqnarray}
The singular perturbation problem in the continuum [whose solution $f(x)$ in
(\ref{BL4}) does not possess a Taylor expansion in powers of $\epsilon$],
has
become a {\it regular} perturbation problem. That is, we can now expand the
solution $f_n$ to the difference equation (\ref{BL6}) as a {\it Taylor
series}
in powers of $\delta$:
\begin{eqnarray}
\label{BL7}
f_n=a_{n,0}+a_{n,1}\delta+a_{n,2}\delta^2+\ldots\, .
\end{eqnarray}

We impose the boundary values (\ref{BL6a}) by requiring that
\begin{eqnarray}
\label{BL8}
a_{0,0}\equiv0\quad{\rm and}\quad a_{n,0}\equiv1\quad(n\geq1).
\end{eqnarray}
Inserting the {\it ansatz} (\ref{BL7}) into the difference equation
(\ref{BL6}),
we get the recursion relation \cite{Bender1}
\begin{eqnarray}
\label{BL9}
a_{n,j}=\frac{1}{2}a_{n+1,j-1}+a_{n,j-1}+\frac{1}{2}a_{n-1,j-1}-\sum_{k=1}^{
j-1}
a_{n,k} a_{n,j-k}-\frac{1}{2}\sum_{k=1}^{j-1}\sum_{l=1}^{j-k}a_{n,k}a_{n,l}
a_{n,j-k-l}.
\end{eqnarray}
For the first derivative at the origin $x=0$ this leads to the series
\begin{eqnarray}
\label{BL10}
f'(0)&=&\lim_{a\to0}\frac{f_1-f_0}{a}=\lim_{a\to0}\frac{f_1}{a}=\lim_{a\to0}
\frac{1}{a}\sum_{j=0}^{\infty}a_{1,j}\delta^j\nonumber\\
&=&\lim_{a\to0}\frac{1}{a}\left(1-\frac{\delta}{2}+\frac{\delta^2}{8}+
\frac{11 \delta^4}{128}+ ... \right).
\end{eqnarray}

We have calculated the coefficients $a_{n,j}$ with the help of Maple V R7 up
to
order $j=200$. The first 20 numbers are given in Table \ref{tab1}. A
complete
list of these coefficients can be found on the webpage of the author FW
\cite{internet2}. Note that the expansion parameter $\delta$ in (\ref{BL10})
is
not small but rather tends to infinity in the limit as the lattice spacing
$a$
approaches zero. Using the parameter $\delta$ defined in (\ref{delta}) we
rewrite the series (\ref{BL10}) as
\begin{eqnarray}
\label{BL12}
f'(0)=\frac{1}{\epsilon}\lim_{\delta\to\infty}\sqrt{\delta}\left(1-
\frac{\delta}{2}+\frac{\delta^2}{8}+\frac{11 \delta^4}{128}+ ... \right).
\end{eqnarray}
Taking into account the exact result (\ref{BL4a}), we obtain the identity
\begin{eqnarray}
\label{BL11}
\frac{1}{\sqrt{2}}=\lim_{\delta\to\infty}\sqrt{\delta}\left(1-\frac{\delta}{
2}+
\frac{\delta^2}{8}+\frac{11\delta^4}{128}+ ... \right).
\end{eqnarray}

The purpose of this paper is to examine equations like (\ref{BL11}). This
equation shows that the singular nature of the instanton problem has
resurfaced
in the continuum limit $\delta\to\infty$ of the lattice expansion. The
expression on the right side of (\ref{BL11}) should have the value
$1/\sqrt{2}=0.7071067812 \ldots$, but it is not at all obvious why this is
so,
and the objective of this paper is to analyze this difficult and subtle limit.

This paper is organized as follows. In Sec.~\ref{sec2} we use Pad\'e
techniques
to perform the limit in (\ref{BL11}). We will see that while the results are
not
bad (the accuracy is about 1\%), better methods are needed. We perform the
Pad\'e analysis to much higher order than has ever been done before and we
discover a new qualitative behavior that has not yet been observed. In
Sec.~\ref{sec3} we try the use of the variational perturbation theory
techniques
introduced by Kleinert to perform the sum in (\ref{BL11}). These techniques
increase the accuracy by a factor of about $10$, but they still do not give
the
exact result. While variational perturbation theory works very well in
summing
the strong-coupling series for the ground-state energy of the
anharmonic oscillator \cite{trick}, and for the critical exponents of
second-order phase transitions \cite{Frohlinde}, we show that the
series in (\ref{BL11}) is at the very edge of validity for Kleinert's
methods.
We then examine the large-order behavior of the terms of the sum in
(\ref{BL11})
in Sec.~\ref{sec4}. We show definitively that the Taylor expansion has a
nonzero
radius of convergence and thus, on the lattice, the instanton problem is a
regular perturbation problem.

In Sec.~\ref{sec5} we turn to a more difficult singular perturbation
problem;
namely, the Blasius equation of fluid dynamics. We use the same approach as
for
the instanton equation. In Secs.~\ref{sec6}, \ref{sec7}, and \ref{sec8} we
study
the summation of the lattice perturbation expansion using Pad\'e and
variational
methods and we examine the large-order behavior of the lattice perturbation
series. We find that Pad\'e methods give good but not excellent results and
that
variational perturbation theory is better than Pad\'e. Again, the series we
need to evaluate in the continuum limit lies at the very edge of validity
for
Kleinert's methods. We also find that, unlike the lattice perturbation
expansion
coefficients for the instanton problem, the sign pattern of the Blasius
weak-coupling series does not alternate. Rather, it is governed by a cosine
function with a frequency different from $\pi$.

\typeout{==> Section 2}
\section{Pad\'e Resummation for the Instanton Equation}
\label{sec2}

In this section we examine what happens if we attempt to evaluate the right
side of (\ref{BL11}) by using Pad\'e techniques. Pad\'e resummation has
already
been applied to the instanton problem up to 50th order \cite{Bender1}.
However,
we have been able to perform the procedures to much higher orders. We have
discovered that remarkable and unsuspected new phenomena occur just a few
orders
beyond what has been computed.

The procedure is as follows. Consider the formal Frobenius series
\begin{eqnarray}
\label{BL13}
S(\delta)=\delta^M\sum_{n=0}^\infty a_n\delta^n,
\end{eqnarray}
where $M$ is a non-negative number. Raising this series to the power $1/M$,
inverting the right hand side and re-expanding, we obtain
\begin{eqnarray}
\label{BL14}
S^{1/M}(\delta)=\frac{\delta}{\displaystyle \sum_{n=0}^\infty b_n\delta^n},
\end{eqnarray}
with new expansion coefficients $b_n$. Assuming we know the first $N+1$
terms
of the original power series in (\ref{BL13}), we raise equation (\ref{BL14})
to
the power $N$. We then truncate the summation at $n=N$, finally getting
\begin{eqnarray}
\label{BL15}
S^{N/M}(\delta)=\frac{\delta^N}{\displaystyle\sum_{n=0}^N
c_n^{(N)}\delta^n},
\end{eqnarray}
where we have re-expanded and obtained new expansion coefficients $c_n$. In
the
limit $\delta\to\infty$, only the $N$th term in the denominator survives and
we
obtain the approximant
\begin{eqnarray}
\label{BL16}
(S_N)^{N/M}\equiv\lim_{\delta\to\infty}S^{N/M}(\delta)=\lim_{\delta\to\infty
}
\frac{\delta^N}{\displaystyle\sum_{n=0}^Nc_n^{(N)}\delta^n}=\frac{1}{c_N^{(N
)}}.
\end{eqnarray}

The approximant $S_N=\left(c_N^{(N)}\right)^{-M/N}$ is the zeroth-order
survivor
of the limiting process. Also, taking into account the first-order
correction we
observe that, as in the case of variational perturbation theory (see
Sec.~\ref{sec3}), there is an approach to scaling. In the limit $\delta\to
\infty$ the Frobenius series $S(\delta)$ in Eq.~(\ref{BL13}) converges to a
constant $C$. Additionally, the approach to scaling, following from the
Pad\'e
resummation (\ref{BL16}), reveals how fast it converges:
\begin{eqnarray}
\label{BL16a}
S(\delta)\sim C +C'\delta^{-1} \quad(\delta\to\infty).
\end{eqnarray}

We now apply this procedure to the boundary-layer problem (\ref{BL6}).
[Recall
that the weak-coupling coefficients for the first 20 coefficients $a_{1,j}$
obtained from (\ref{BL9}) are shown in Table \ref{tab1} and that more can be
found in \cite{internet2}.] Resumming the series (\ref{BL7}) for $n=1$,
\begin{eqnarray}
\label{BL17}
f_1=\sum_{j=0}^N a_{1,j}\delta^j,
\end{eqnarray}
according to the Pad\'e procedure (\ref{BL16}) with $M=1/2$ as follows from
(\ref{BL11}) and evaluating the approximants $S_N=\left(c_N^{(N)}\right)^{-
M/N}$, we get the numbers listed in Table \ref{tab}.

\begin{figure}[htbp]
\centerline{\epsfysize=10cm \epsfbox{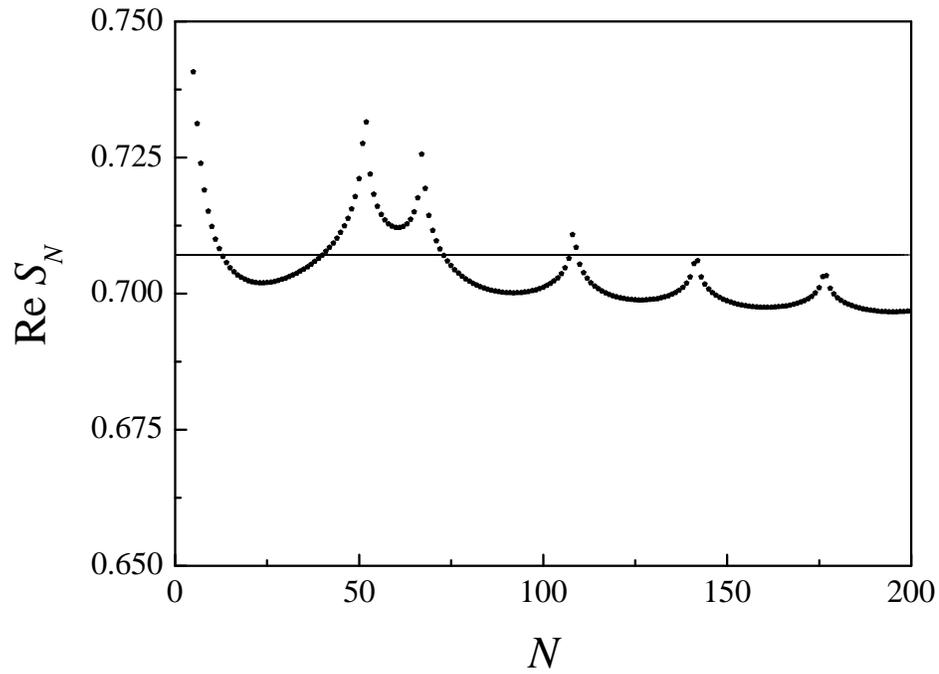}} 
\caption[The approximants for the instanton problem up to 200th order]
{\label{strong}
The real part of the Pad\'e approximants $S_N$ up to 200th order. Note that
the approximants do not converge to the exact solution, which is represented
by
the horizontal solid line. The phases where the approximants become complex
are
marked by spikes.}
\end{figure}

Compared with the numerical solution $1/\sqrt{2}\approx 0.7171067812$, this
strong-coupling expansion seems to converge quite well. However, when we go
to
higher orders, we find that the numbers drop below the exact solution and
assume a minimum at $N=24$, where the approximant has the value
$S_{24}\approx
0.70198319$. The approximants then rise again, cross the exact solution at
$N=41$ and become complex at $N=52$. The appearance of complex numbers is a
consequence of taking the $N$th root in equation (\ref{BL16}) when the
coefficients $c_N^{(N)}$ become negative. This phenomenon has not been
observed
before in the course of using this Pad\'e procedure. The imaginary part then
becomes smaller and smaller as $N$ rises. Abruptly, at $N=68$, the
approximants
become real again. As one can see from the spikes in Fig.~{\ref{strong}}
this
pattern is repeated for higher $N$. Note that the figure only shows the real
part of the Pad\'e approximant $S_N$.

Apparently, the sequence of approximants $S_N$ does not converge. The
singular
nature of the instanton equation has the effect of making the Pad\'e
approximants behave like the partial sums of a divergent (asymptotic)
series; at
first the partial sums appear to converge to a limit, and then they veer
off. In
the case of the Pad\'e's shown in Fig.~\ref{strong} the approximants
approach to
within 1\% of the correct limit before veering off. It appears that another
more powerful resummation technique is needed to treat the expression in
(\ref{BL11}). In the next section we apply a technique due to Kleinert.

\typeout{==> Section 3}
\section{Variational Perturbation Theory for the Instanton Equation}
\label{sec3}

Kleinert has developed a technique in the context of the ground-state energy
of
the anharmonic oscillator \cite{trick} and of critical exponents of
second-order
phase transitions \cite{Frohlinde} for summing divergent perturbation
series.
This technique, known as Kleinert's square-root trick, is described below.

Consider a weak-coupling series
\begin{eqnarray}
\label{VPT1}
f_N(\delta)=\sum_{n=0}^{N} f_n \delta^n,
\end{eqnarray}
which is truncated at order $N$. Rewrite this weak-coupling expansion by
introducing an auxiliary scaling parameter $\kappa$ \cite{trick,Frohlinde}:
\begin{eqnarray}
\label{VPT2}
f_N(\delta)=\kappa^p \sum_{n=0}^{N} f_n \left( \frac{\delta}
{\kappa^q} \right)^n \Big|_{\kappa=1},
\end{eqnarray}
which is set to $\kappa=1$ later.
The square-root trick now reads
\begin{eqnarray}
\label{VPT3}
\kappa \rightarrow \sqrt{K^2 +\kappa^2-K^2}=K\sqrt{1+\delta r},
\end{eqnarray}
where $K$ is a ``dummy'' scaling parameter and
\begin{eqnarray}
\label{VPT3b}
r=\frac{1}{\delta}\left(\frac{\kappa^2}{K^2}-1 \right).
\end{eqnarray}
In the case of the anharmonic oscillator, $K$ is the frequency $\Omega$ of a
trial harmonic oscillator \cite{trick}.

Substituting (\ref{VPT3}) into the truncated weak-coupling series
(\ref{VPT2}),
we obtain
\begin{eqnarray}
\label{VPT3c}
f_N(\delta,K) =\sum_{n=0}^N f_n K^{p-nq}(1+\delta r)^{(p-nq)/2}\delta^n \, .
\end{eqnarray}
The factor $(1+gr)^{\alpha}$ with $\alpha\equiv(p-nq)/2$ can be expanded by
means of generalized binomials according to
\begin{eqnarray}
\label{VPT3d}
(1+\delta r)^{\alpha}=\sum_{k=0}^{N-n}
{\alpha \choose k} (\delta r)^k \delta^n
=\sum_{k=0}^{N-n} {\alpha \choose k} \left( \frac{1}{K^2}-1\right)^k \delta ^n \,
,
\end{eqnarray}
where we have used (\ref{VPT3b}) and finally have set $\kappa \equiv 1$. The
binomial is defined as
\begin{eqnarray}
\label{VPT3f}
{\alpha \choose k}\equiv
\frac{\Gamma (\alpha+1)}{\Gamma(k+1)\Gamma(\alpha+k+1)} \, .
\end{eqnarray}

We deduce that the function $f_N(\delta,K)$ can now be written as
\begin{eqnarray}
\label{VPT4}
f_N(\delta,K)=\sum_{n=0}^N \left[ \sum_{k=0}^{N-n} {\frac{1}{2}(p-nq) \choose k}
\left( \frac{1}{K^2}-1 \right)^k K^{p-nq} \right] f_n \delta^n \, .
\end{eqnarray}
To first order this expression reduces to
\begin{eqnarray}
\label{VPT4c}
f_1(\delta,K) = \left( 1-\frac{p}{2} \right) f_0 K^p +\frac{p}{2} f_0 K^{p-2}
+f_1 \delta K^{p-q} \, .
\end{eqnarray}
Applying the principle of least sensitivity \cite{Stevenson} leaves us with
\begin{eqnarray}
\label{VPT4d}
\frac{\partial f_1(\delta,K)}{\partial K} \sim p \left( 1-\frac{p}{2} \right) f_0
+ \frac{p(p-2)}{2}f_0 K^{-2} +(p-q) f_1 \delta K^{-q} \equiv 0 \, .
\end{eqnarray}

Next, making the strong-coupling ansatz
\begin{eqnarray}
\label{sahne2}
K^{(1)}(\delta)=\delta^{1/q} \left( k_0^{(1)}+k_1^{(1)} \delta^{-2/q} + ... \right) \, ,
\end{eqnarray}
we obtain the following equation from (\ref{VPT4d}):
\begin{eqnarray}
\label{sahne3}
p \left( 1-\frac{p}{2} \right) f_0+\frac{p(p-2)}{2} f_0 (k_0^{(1)}
\delta^{1/q})^{-2}
+(p-q) f_1 \delta (\delta^{1/q} k_0^{(1)})^{-q} = 0\, .
\end{eqnarray}
The second term is a subleading contribution in the limit as
the coupling $\delta$ goes to infinity which we can neglect.
Solving for $k_0^{(1)}$ we then get
\begin{eqnarray}
\label{sahne4}
k_0^{(1)}= \left( \frac{2 f_1}{f_0} \frac{p-q}{p(p-2)} \right)^{1/q} \, .
\end{eqnarray}
Assuming that the ansatz (\ref{sahne2}) for the variational parameter $K(\delta)$
also holds for higher orders we obtain from the function $f_N(\delta,K)$ in
(\ref{VPT4})
\begin{eqnarray}
\label{VPT5}
f_N(\delta)=\delta^{\frac{p}{q}}
\left[ b_0^{(N)}(k_0^{(N)})+b_1^{(N)}(k_0^{(N)}, k_1^{(N)})
\delta^{-2/q}+... \right] \, ,
\end{eqnarray}
where the leading strong-coupling coefficient $b_0^{(N)}(k_0^{(N)})$ is
given by
\begin{eqnarray}
\label{sahne5}
b_0^{(N)}(k_0^{(N)}) =\sum_{n=0}^{N} \sum_{k=0}^{N-n}
{\frac{1}{2} (p-nq) \choose k}(-1)^k f_n (k_0^{(N)})^{p-nq}\,.
\end{eqnarray}

The inner sum can be further simplified, using
\begin{eqnarray}
\label{VPT5b}
\sum_{k=0}^m (-1)^k {\alpha \choose k}=(-1)^m{\alpha -1\choose m} \, .
\end{eqnarray}
Thus the strong-coupling coefficient (\ref{sahne5}) reduces to
\begin{eqnarray}
\label{BL18}
b_0^{(N)}(k_0^{(N)})=\sum_{n=0}^N (-1)^{N-n} {\frac{1}{2} (p-nq)-1 \choose
N-n}
f_n (k_0^{(N)})^{p-nq} \, .
\end{eqnarray}
So, looking at equation (\ref{VPT5}) we see that the fraction $p/q$ tells us
the
leading power behavior in $\delta$ and $2/q$ indicates the approach to scaling:
\begin{eqnarray}
\label{allgemein}
\sum_{j=0}^\infty f_j\delta^j\sim\delta^{p/q}\left(b_0+b_1\delta^{-2/q}+ ...
\right)\quad(\delta\to\infty).
\end{eqnarray}

For the instanton equation we can determine the numbers $p$ and $q$ by
re-obtaining the differential equation (\ref{BL3}) from the difference
equation
(\ref{BL6}). The positive real axis is discretized in steps of width $a$, so
that we let $x_n\equiv na$. The power series expansion for the discrete
function
$f_n=f(x_n)$ has the form
\begin{eqnarray}
\label{T1}
f_{n\pm 1}=f(x_n)\pm
f'(x_n)a+\frac{1}{2}f''(x_n)a^2\pm\frac{1}{6}f'''(x_n)a^3
+\frac{1}{24}f''''(x_n)a^4\pm ...\, .
\end{eqnarray}
Thus, the numerator of the second derivative (\ref{BL5}) becomes
\begin{eqnarray}
\label{T2}
f_{n+1}-2f_n+f_{n-1}=f''_n a^2+\frac{1}{12}f''''_n a^4+ ... \, ,
\end{eqnarray}
so the zeroth-, first-, and third-order contributions cancel. Translating
the
lattice result for $f_n$ back to the continuous function $f(x_n)=f_n$, the
difference equation (\ref{BL6}) reads
\begin{eqnarray}
\label{T3}
\epsilon^2\left[f''(x)+\frac{1}{12}f''''(x)a^2+ ...\right]+f(x)-f^3(x)=0.
\end{eqnarray}

Writing out the power series
\begin{eqnarray}
\label{T4}
f(x)=f_0(x)+a^2f_1(x)+a^4f_2(x)+ ...,
\end{eqnarray}
and comparing even powers of $a$, we get from equation (\ref{T3}) for $a^0$
\begin{eqnarray}
\label{T5}
\epsilon^2 f''_0(x)+f_0(x)-f_0^3(x)=0,
\end{eqnarray}
which is just the original instanton equation (\ref{BL3}). For $a^2$ we have
\begin{eqnarray}
\label{T6}
\epsilon^2 f''_1(x)+f_1(x)\left(1-3f_0^2(x)\right)=-\frac{1}{12}\epsilon^2
f''''_0(x).
\end{eqnarray}

The boundary values read
\begin{eqnarray}
\label{mist}
f_0(0)=0, \qquad f_0(\infty)=1,
\end{eqnarray}
and
\begin{eqnarray}
f_1(0)=f_1(\infty)=0,
\end{eqnarray}
respectively. The solution to equation (\ref{T5}) with the boundary values
(\ref{mist}) is of course
\begin{eqnarray}
\label{T6b}
f_0(x)=\tanh\frac{x}{\epsilon\sqrt{2}}.
\end{eqnarray}
So, finally from (\ref{T4}) we get for the derivative at the origin $x=0$:
\begin{eqnarray}
\label{T7}
f'(0)=f_0'(0)+\frac{\epsilon^2}{\delta}f_1'(0)+ ... =
\frac{1}{\epsilon \sqrt{2}}+\frac{\epsilon^2}{\delta}f'_1(0) + ...\,.
\end{eqnarray}

Comparing equation (\ref{T7}) with (\ref{BL12}), we resum the weak-coupling
series in (\ref{BL12}) as
\begin{eqnarray}
\label{T8}
1-\frac{\delta}{2}+\frac{\delta^2}{8}+
...=\delta^{-1/2}\left[\frac{1}{\sqrt{2}}
+\epsilon^3 f_1'(0)\delta^{-1}+ ...\right].
\end{eqnarray}
Also, comparing with (\ref{allgemein}), we conclude that the leading power
and
the approach to scaling are given by
\begin{eqnarray}
\label{T9}
\frac{p}{q}=-\frac{1}{2},\qquad\frac{2}{q}=1,
\end{eqnarray}
respectively. So we identify $p=-1$ and $q=2$.

We now evaluate the leading strong-coupling coefficient $b_0$ from
(\ref{allgemein}) according to (\ref{BL18}) with $p=-1$ and $q=2$. To that
end
we substitute our 200 weak-coupling coefficients from \cite{internet2}
into the formula using a
computer algebra program. We are now confronted with the following problem:
The principle of least sensitivity cannot be unambiguously applied.
Optimizing
with respect to extrema, inflection points, or higher derivatives does yield
converging results for the strong-coupling limit. However, all these
strong-coupling series converge to the wrong values.

\begin{figure}[htbp!]
\centerline{\epsfysize=10cm \epsfbox{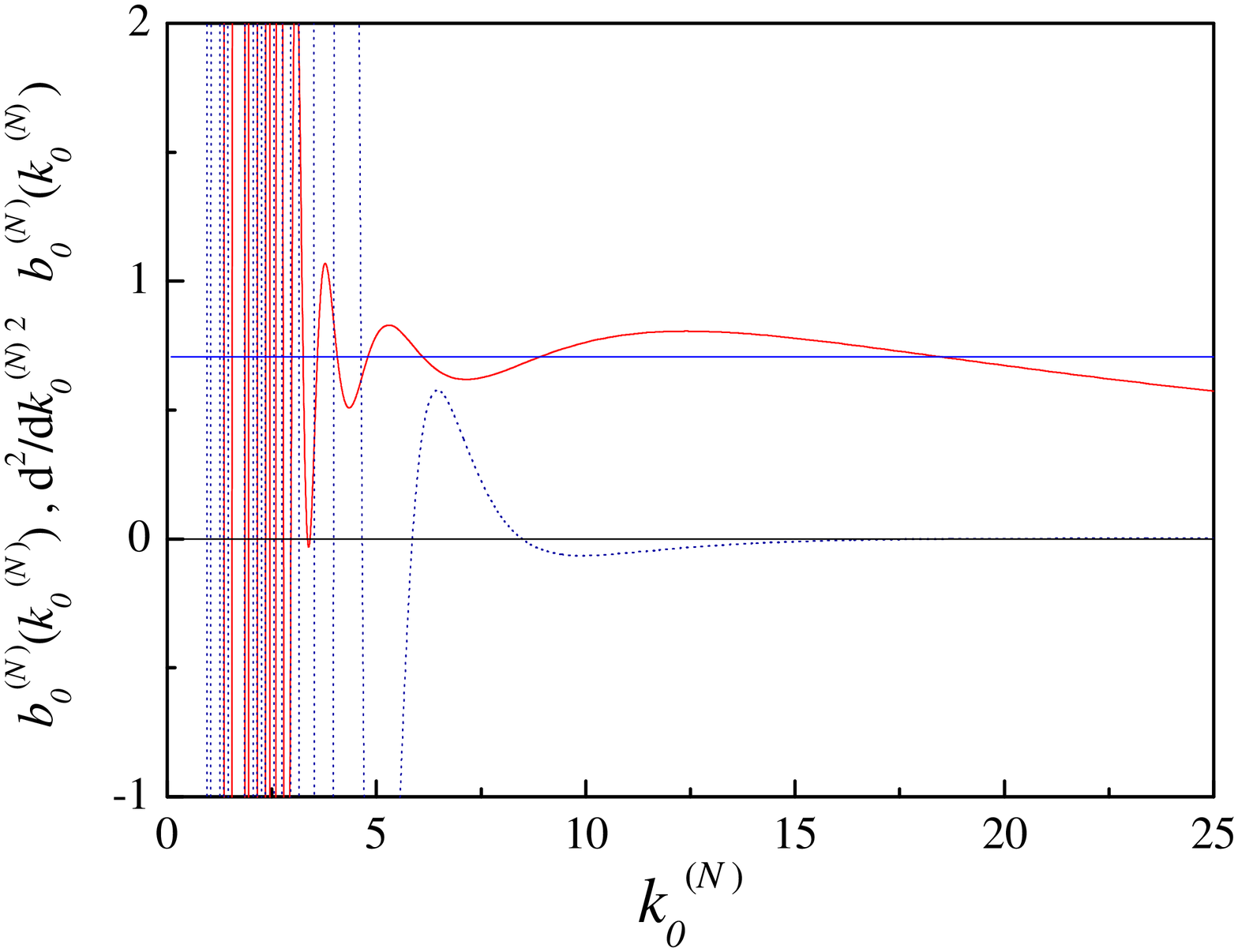}} 
\caption[The function $b_0^{(200)}(k_0)$ and its second derivative with
respect
to $k_0$, the variational parameter in the case of the instanton problem.]
{\label{VPTpic} The function $b_0^{(N)}(k_0^{(N)})$ from (\ref{BL18})
for $N=200$
(solid line) and its second derivative with respect to $k_0$ (dotted line).
The
upper horizontal line equals $1/\sqrt{2}$, the correct limiting value of the
instanton problem. All extrema of $b_0^{(N)}$ are far from this value. Only
the
inflection point on the right-hand side comes close. The value for $k_0$,
for
which the second derivative vanishes, is $k_0=18.42510$. Substituting that
number into the function $b_0^{(N)}(k_0^{(N)})$, we obtain in the 200th
order
$b_0^{(200)}=0.707417$. The corresponding Richardson extrapolations
can be found in Table \ref{VPTtab}.}
\end{figure}

There is one particularly unpleasant case: The second derivative with
respect to
$k_0$ for the largest $k_0$ where this derivative exists (see
Fig.~\ref{VPTpic})
gives a convergent strong-coupling series. The numbers come extremely close
to
$1/\sqrt{2}$ as one can see from the 20 numbers in Table \ref{VPTapp}. The
200th
leading strong-coupling coefficient is $b_0^{(200)}=0.707417...$. However, a
Richardson extrapolation \cite{BO} based on the first 200 orders then
unfortunately shows that variational perturbation theory produces a value
slightly smaller than $1/\sqrt{2}$. The first six orders of Richardson
extrapolations are presented in Table \ref{VPTtab}. Hence, the
strong-coupling
series $b_0^{(N)}$ does converge, but it converges to the wrong number, only
one
part per 1000 away from the true value:
\begin{eqnarray}
\label{konstantin}
f_1^{\rm (VPT)} \approx
\lim_{\delta\to\infty}\sum_{n=0}^{200}a_{1,n}\delta^n
= b_0^{(\infty)}=0.7063998320858845\pm0.0000000000000001\,
\end{eqnarray}
compared with $f'(0)=1/\sqrt{2}=0.7071067812...$~. The deviation is just
$0.099\%$, but $1/\sqrt{2}$ can unfortunately be ruled out.

Given that $p=-1$ and $q=2$, the failure of variational perturbation theory
is
not surprising. According to Ref.~\cite{Frohlinde} the fraction $2/q$ must
lie
within the open interval $(1/2,1)$. Otherwise, one cannot prove that
variational
perturbation theory converges. Thus, {\it this problem lies exactly on the
upper boundary of the region in which the summation method is known to work}.

We can understand the upper edge of the range of the parameter $2/q$ that
describes the approach to scaling $2/q$ by looking at the standard deviation
from the actual limiting value. It turns out \cite{Frohlinde} that the
deviation
in the limit as the perturbative order $N$ goes to infinity assumes the
shape
\begin{eqnarray}
\label{T10}
\left|\frac{b_0^{(N)}-b_0}{b_0}\right|\sim
\exp\left(-CN^{1-2/q}\right) \quad (N\to\infty),
\end{eqnarray}
where $C$ is a constant. So, to obtain exponential convergence for the
sequence
formed by the $b_0^{(N)}$, we need $1-2/q>0$. In other words, the approach
to
scaling $2/q$ is bounded and it must be smaller than one. The lower edge is
more
subtle and is discussed in Ref.~\cite{Frohlinde}.

In conclusion, we have applied variational perturbation theory to a case
that
lies at the very edge of its applicability. We see that variational
perturbation
theory gives better results by about a factor of 10 than the Pad\'e
approximations examined in Sec.~\ref{sec2}. However, we have not yet found a
systematic method for resumming (\ref{BL11}) that enables us to perform the
continuum limit of the discrete lattice theory. Therefore, we now lay the
foundation for further investigations by analyzing the large-order behavior
of
the instanton series.

\typeout{==> Section 4}
\section{Large-Order Behavior for the Instanton Equation}
\label{sec4}

It can be seen from the numerical results in \cite{internet2} that the
instanton
weak-coupling series is of Borel type. That is, it exhibits an alternating
sign
pattern. From the ratio test we can see that the coefficients $a_{n,j}$ do
not
grow factorially fast. The large-order behavior of $a_{n,j}$ has the general
form
\begin{eqnarray}
\label{flori1}
a_{n,j}\sim(-1)^{n+j+1}K_n^j j^{A_n}B_n\quad (j\to\infty).
\end{eqnarray}
The constant $A_n$ can be obtained by evaluating the limit
\begin{eqnarray}
\label{flori2}
A_n=\lim_{j\to\infty}\frac{\log\displaystyle\frac{a_{n,j+2}\,a_{n,j}}{
(a_{n,j+1})^2}}{\log \displaystyle \frac{j(j+2)}{(j+1)^2}},
\end{eqnarray}
and the reciprocal of the radius of convergence is
\begin{eqnarray}
\label{flori3}
K_n=-\lim_{j\to\infty}\frac{a_{n,j+1}}{a_{n,j}}\left(\frac{j}{j+1}\right)^{A
_n}.
\end{eqnarray}
Also, the overall factor $B_n$ is determined from
\begin{eqnarray}
\label{flori4}
B_n=\lim_{j\to\infty}\frac{|a_{n,j}|}{K_n^j j^{A_n}}.
\end{eqnarray}

Using the 200 weak-coupling coefficients, we find that the exponent $A_n$
and
the reciprocal radius of convergence $K_n$ are independent of $n$. The value
of
$K_2=2.46682906$ coincides with $K_1=2.46682906$ for all significant digits.
The
same is true for $A_1=-1.500000$ and $A_2=-1.500000$. Thus, it appears that
we
may omit the subscripts $n$ for $K_n$ and $A_n$.
In contrast, the data suggests that $B_n$ strongly
depends on $n$. $B_n$ is the numerical value associated with the largest
uncertainty. In fact, Eq.~(\ref{flori4}) suggests that small deviations in
$K$
and $A$ lead to dramatic changes in the value of $B_n$.
We calculated $A$, $K$, $B_1$, and $B_2$ up to 200th order
with the help of Maple V R7. We then extrapolated these 200 orders to
infinity
using Richardson extrapolation \cite{BO}. We obtained
\begin{eqnarray}
\label{flori5}
A&=&-1.500000\pm 0.000001,\nonumber\\
K&=&2.46682906\pm 0.0000001,\nonumber\\
B_1&=&0.0171\pm 0.0001,\nonumber\\
B_2&=&0.1190\pm 0.0001.
\end{eqnarray}

Detailed numerical results for the first six Richardson extrapolations for
the
exponent $A$, the inverse radius of convergence $K$, and the overall factors
$B_1$ and $B_2$ can be found in Tables \ref{richy2}, \ref{richy3},
\ref{richy4}, and \ref{richy5}. The
calculation of $B_1$ is extremely delicate; changing the inverse radius of
convergence in the sixth decimal place influences the third significant
figure
of $B_1$. The same is true of $B_2$.

Unfortunately, there is no way to derive these values by applying asymptotic
analysis to the recursion relation (\ref{BL9}). The problem is that the
double
summation in this equation includes small $j$, so we cannot let $j$ go to
infinity and use the large-order behavior (\ref{flori1}). Substituting the
{\it
ansatz} (\ref{flori1}) into equation (\ref{BL9}) and taking the limit leads
to
contradictory results. For $n=1$ we get
\begin{eqnarray}
\label{flori6}
Kj^A B_1 &=&\frac{1}{2}(j-1)^AB_2+(j-1)^AB_1-\frac{3}{2}B_1^2
K\sum_{k=1}^{j-1}
k^A (j-k)^A\nonumber\\
&& -\frac{1}{2}B_1^3K\sum_{k=1}^{j-1}\sum_{l=1}^{j-k}k^Al^A(j-k-l)^A.
\end{eqnarray}

Pulling out some factors and letting $x\equiv k/j$, we obtain for the first
summation
\begin{eqnarray}
\label{flori8}
\lim_{j\to\infty}\sum_{k=1}^j\left(\frac{k}{j}\right)^A\left(1-\frac{k}{j}
\right)^A=\int_0^1 dx\,[x(1-x)]^A=\frac{\Gamma^2(A+1)}{\Gamma(2A+2)},
\end{eqnarray}
if and only if $A>-1$. For $A<-1$ which is strongly favored by
the data we obtain
\begin{eqnarray}
\label{flori8a}
\int_0^1 dx\,[x(1-x)]^A=2\zeta\left(-A\right).
\end{eqnarray}

The double summation reduces to
\begin{eqnarray}
\label{flori9}
\lim_{j\to\infty}\sum_{k=1}^j\sum_{l=1}^{j-k}\frac{k^A l^A}{j^{2A}} \left(1
-\frac{k}{j}-\frac{l}{j}\right)=\int_0^1 dx\,\int_0^1 dy\,[xy(1-x-y)]^A=
\frac{\Gamma^3(A+1)}{\Gamma(3A+3)},
\end{eqnarray}
where $y\equiv l/j$ and $A>-1$. For $A<-1$ the result is
\begin{eqnarray}
\label{flori9a}
\int_0^1 dx\,\int_0^1 dy\,[xy(1-x-y)]^A=3\zeta^2\left(-A\right).
\end{eqnarray}

Substituting the results in (\ref{flori8a}) and (\ref{flori9a}) into
(\ref{flori6}) leads to a contradiction: The inverse radius of convergence
then
turns out to be
\begin{eqnarray}
\label{AAA}
K=\frac{1+\frac{B_2}{2B_1}}{1+3\zeta\left(\frac{3}{2}\right)B_1+\frac{3}{2}
\zeta^2\left(\frac{3}{2}\right)B_1^2},
\end{eqnarray}
which would imply that, given $B_1=0.0171$ and $B_2=0.1190$, the value of
$K$
would be
\begin{eqnarray}
\label{AAA2}
K=3.940.
\end{eqnarray}
This result can be ruled out because of the numerical result (\ref{flori5}).
Also, (\ref{AAA}) does not contain the exponent $A$ because all the factors
$j^A$ in (\ref{flori6}) cancel. So $A$ cannot be determined analytically
using
this asymptotic analysis.

\typeout{==> Section 5}
\section{Boundary-Layers on the Lattice --- Blasius Equation}
\label{sec5}

The Blasius equation \cite{blasius} arises in the study of fluid dynamics.
It is a special limiting case of the Navier-Stokes equation and determines
the flow of an incompressible fluid across a semi-infinite flat plate.
The equation reads
\begin{eqnarray}
\label{blasius1}
2\epsilon y'''(x)+y(x)y''(x)=0.
\end{eqnarray}
Assuming that the tangential velocity $y'(x)$ at the outer limit of the
boundary layer is constant, the boundary conditions read \cite{acta}
\begin{eqnarray}
\label{blasius2}
y(0)=y'(0)=0, \qquad y'(\infty)=1.
\end{eqnarray}

Our objective here is to calculate the second derivative $y''(0)$, which
represents the stress on the plate. We discretize the Blasius equation
(\ref{blasius1}) by introducing a lattice spacing $a$:
\begin{eqnarray}
\label{blasius3}
2\delta(f_{n+1}-3f_n+3f_{n-1}-f_{n-2})+f_n(f_{n+1}-2f_n+f_{n-1})=0,
\end{eqnarray}
where we define $f_n\equiv y(na)/a$ and $\delta\equiv\epsilon/a^2$.
The boundary conditions (\ref{blasius2}) now read
\begin{eqnarray}
\label{blasius3a}
f_0=f_{-1}=0, \qquad f_n\sim n\quad(n\to\infty).
\end{eqnarray}

Expanding $f_n$ as a series in powers of $\delta$ as in Eq.~(\ref{BL7}), we
obtain the recursion relation \cite{Bender1}
\begin{eqnarray}
\label{blasius4}
a_{n+1,j}-2a_{n,j}+a_{n-1,j}&=& -\frac{2}{n}\left(a_{n+1,j-1}-3a_{n,j-1}
+3a_{n-1,j-1}-a_{n-2,j-1}\right)\nonumber\\
&&-\frac{1}{n}\sum_{k=1}^{j-1}a_{n,k}\left(a_{n+1,j-k}-2a_{n,j-k}+a_{n-1,j-k
}
\right),
\end{eqnarray}
The boundary values are
\begin{eqnarray}
\label{hmmm}
a_{n,0}&=&n ~ (n\geq0),\nonumber\\
a_{-1,0}&=&0,\nonumber\\
a_{-n-1,j}&=&a_{n,j}~(n\geq0).
\end{eqnarray}
Eq.~(\ref{blasius4})
can be solved order by order by using a computer algebra program. Table
\ref{tabelle} shows the first 20 weak-coupling coefficients $a_{1,j}$. All
coefficients up to the 300th order can be found at \cite{internet3}.

\typeout{==> Section 6}
\section{Pad\'e Resummation for the Blasius Equation}
\label{sec6}

We now resum the weak-coupling coefficients using the Pad\'e method
(\ref{BL16})
with $M=-1/2$. This value of $M$ will be derived in Sec.~\ref{sec7} in
Eq.~(\ref{lory7}). The exact solution \cite{Bender1} to the Blasius equation
(\ref{blasius1}), obtained numerically up to five digits, is
$y''(0)=0.33206$.
Unfortunately, the sequence formed by the approximants $S_N$ appears to
converge, but not to the correct value. According to Table \ref{tabelle2}
the
sequence becomes very flat and Richardson extrapolation \cite{BO} shows that
the $S_N$ approach the wrong limiting value (see Table \ref{richy}). A
third-order Richardson gives $S_{\infty}=0.3430$, based on the first 70
weak-coupling coefficients. This value is significantly higher than the
correct
value $y''(0)=0.33206$, the deviation is $3.3\%$.

The failure of the Pad\'e resummation is not surprising because the Pad\'e
method assumes the approach to scaling $\delta^{-1}$ according to
(\ref{BL16a}).
However, in the case of the Blasius equation the approach to scaling is
$\delta^{-1/2}$, as we will see in equation (\ref{lory7}) in the next
section.

\typeout{==> Section 7}
\section{Variational Perturbation Theory for the Blasius Equation}
\label{sec7}

Variational perturbation theory for the Blasius equation fails to converge
to
the correct answer in the same way as for the instanton problem. We
determined
the leading strong-coupling term (\ref{BL18}) up to 200th order
and again it was
impossible to find extrema, inflection points, or higher derivatives that
yield the correct result. Tables \ref{bald} and \ref{bald2} show the last 20
strong-coupling coefficients $b_0^{(N)}$ and six orders of Richardson
extrapolation.
By determining the values of $p$ and $q$ we show why
variational perturbation is likely to fail for this problem.

Consider again the Taylor expansions for $f_{n \pm 1}$ in (\ref{T1})
together
with the Taylor series for $f_{n-2}=f(x_n-2a)$, namely
\begin{eqnarray}
\label{lory1}
f_{n-2}=f(x_n)-2f'(x_n)a+2f''(x_n)a^2-\frac{4}{3}f'''(x_n)a^3+\frac{2}{3}
f''''(x_n)a^4\pm ...\,.
\end{eqnarray}
Inserting these expressions into the difference equation for the Blasius
problem
(\ref{blasius3}) and translating back to the continuous function
$f(x_n)=f_n$,
we get
\begin{eqnarray}
\label{lory2}
2\epsilon\left(f'''(x)a-\frac{1}{2}f''''(x)a^2+...\right)+f(x)\left(f''(x)a^
2
+\frac{1}{2}f''''(x)a^4 +...\right)=0.
\end{eqnarray}

Next we transform back to the function $y(x)=af(x)$ and assume the Taylor
series
\begin{eqnarray}
\label{lory3}
y(x)=y_0(x)+ay_1(x)+a^2 y_2(x)+ ... \, .
\end{eqnarray}
To zeroth order in $a$ we obtain
\begin{eqnarray}
\label{lory4}
2\epsilon y_0'''(x)+y_0(x)y_0''(x)=0,
\end{eqnarray}
which is just the Blasius equation (\ref{blasius1}). The small parameter
$a$,
which is the lattice spacing, relates $\epsilon$ and $\delta$ by $a=\sqrt{
\epsilon/\delta}$. Thus, if we evaluate the Taylor series (\ref{lory3}) for
the
second derivative at the origin, we see that
\begin{eqnarray}
\label{lory5}
y''(0)=y_0''(0)+ay_1''(0)+ ... =\frac{0.33206}{\sqrt{\epsilon}}+
\sqrt{\frac{\epsilon}{\delta}} y_1''(0)+ ...\,.
\end{eqnarray}

Comparing this series to the original weak-coupling series
\begin{eqnarray}
\label{lory6}
y''(0)=\sqrt{\frac{\delta}{\epsilon}}\left(1-2\delta+2\delta^2+...\right),
\end{eqnarray}
we can now determine the leading power $p/q$ and the approach to scaling
$2/q$:
\begin{eqnarray}
\label{lory7}
1-2\delta+2\delta^2+ ... = \delta^{-1/2}\left(0.33206+\delta^{-1/2}\epsilon
y_1''(0)+ ... \right),
\end{eqnarray}
so we obtain $p=-2$ and $q=4$.

Again we find that the approach to scaling $2/q=1/2$ lies just on the
boundary
of the open interval $(1/2,1)$, for which the proof of convergence
\cite{Frohlinde} holds. This situation here is the opposite of the instanton
case in that it sits at the {\it lower} boundary of the open interval in
which
variational perturbation theory works.

\typeout{==> Section 8}
\section{Large-order Behavior for the Blasius Equation}
\label{sec8}

The Blasius equation exhibits a large-order behavior which is a more subtle
than for the instanton problem (\ref{flori1}). The Blasius weak-coupling
coefficients are not of Borel type; that is, the sign pattern is not
alternating. Rather, the sign structure is governed by a cosine function
with a
frequency that is significantly different from $\pi$. Remarkably, it turns
out
that a pure cosine $\cos(an)$ cannot reproduce all signs correctly. Up to 300th
order the sign structure given by $\cos(an)$ is broken twice: The signs at
$n=62$ and at $n=212$ are not correct if we optimize with respect to $a$.
So we must consider an additional phase
shift $\cos(an+b)$. The parameter $b$ turns out to be slightly smaller than
$\pi$, but it reproduces all 300 signs correctly.

In order to determine the numerical values of $a$ and $b$ we define
\begin{eqnarray}
\label{AB}
f(a,b)\equiv\sum_{n=1}^{N}\frac{\cos(an+b)}{|\cos (an+b)|}\frac{a_{1,n}}
{|a_{1,n}|}.
\end{eqnarray}
The sum ends at $N=300$ because this is as high as we can calculate using
Maple;
we know the first 300 weak-coupling coefficients $a_{1,j}$. For the correct
values of $a$ and $b$ the function $f(a,b)$ must be equal to 300. We then
plot
the function $f(a,b)$ over the $a$--$b$ plane and search for peaks. A
careful
study of the peaks yields values for $a$ and $b$ which allow the function
$f(a,b)$ to assume its maximum at 300. These numbers are given in Table
\ref{tabelle3}.

The large-order behavior of the Blasius weak-coupling coefficients (unlike
the
large-order behavior of the instanton coefficients) has an additional
overall
factor $\cos(an+b)$, and we can now see that the remaining structure differs
from the structure of the instanton weak-coupling coefficients. Dividing by
the
cosine, we observe that the coefficients
\begin{eqnarray}
\label{wasnun}
a'_j\equiv\frac{a_{1,j}}{\cos (aj+b)}
\end{eqnarray}
grow factorially fast. Thus, we also divide by $j!$:
\begin{eqnarray}
\label{wasnun2}
b_j\equiv\frac{a_{1,j}}{\cos(aj+b)j!}.
\end{eqnarray}
The coefficients $b_j$ are unstable under a ratio test. That is, the ratio
$b_{j
+1}/b_j$ decreases and then begins to oscillate. This is the inaccuracy that
results from the delicate sign pattern of the first 300 coefficients
$a_{1,j}$.

\typeout{==> Acknowledgments}
\section* {Acknowledgments}
\label{sec9}

CMB is grateful to the U.S.~Department of Energy for financial support.
FW and AP wish to thank Hagen Kleinert for fruitful discussions
on variational perturbation theory.
Moreover AP acknowledges financial support from the German
Research Foundation (DFG) under contract number KON 1823/2001.

\typeout{==> References}

\newpage
\typeout{==> Tables}
\section*{Tables}

\begin{table}[htbp]
\begin{center}
\begin{tabular}{|c|c||c|c|} \hline
$j$ & $a_{1,j}$        & $j$ & $a_{1,j}$ \\[1mm] \hline \hline
\rule[-5pt]{0pt}{20pt}
1 & $-\frac{1}{8}$     & 11  & $-\frac{2887747}{262144}$  \\[1mm] \hline
\rule[-5pt]{0pt}{20pt}
2 & $\frac{1}{8}$      & 12  & $\frac{99392471}{4194304}$ \\[1mm] \hline
\rule[-5pt]{0pt}{20pt}
3 & 0                  & 13  & $-\frac{215798295}{4194304}$ \\[1mm] \hline
\rule[-5pt]{0pt}{20pt}
4 & $\frac{11}{128}$   & 14  & $\frac{3781670831}{33554432}$ \\[1mm] \hline
\rule[-5pt]{0pt}{20pt}
5 & $-\frac{23}{128}$  & 15  & $-\frac{8349041385}{33554432}$ \\[1mm] \hline
\rule[-5pt]{0pt}{20pt}
6 & $\frac{295}{1024}$ & 16  & $\frac{1188129285795}{2147483648}$
\\[1mm]\hline
\rule[-5pt]{0pt}{20pt}
7 & $-\frac{589}{1024}$ & 17 & $-\frac{2659104132291}{2147483648}$
\\[1mm]\hline
\rule[-5pt]{0pt}{20pt}
8 & $\frac{39203}{32768}$ & 18 & $\frac{47890245452569}{17179869184}$
\\[1mm] \hline
\rule[-5pt]{0pt}{20pt}
9 & $-\frac{80723}{32786}$ & 19 & $-\frac{108383753179167}{17179869184}$
\\[1mm] \hline
\rule[-5pt]{0pt}{20pt}
10 & $\frac{1354949}{262144}$& 20  &
$\frac{39433620359113981}{274877906944}$  \\[1mm] \hline
\end{tabular}
\end{center}
\caption[The first 20 weak-coupling coefficients for the instanton problem]
{\label{tab1}
The first 20 weak-coupling coefficients $a_{1,j}$ for the instanton problem
(\ref{BL8}) and (\ref{BL9}).}
\end{table}

\begin{table}[htbp]
\begin{center}
\begin{tabular}{|c|l||c|l|} \hline
$N$ & $S_N$        & $N$ & $S_N$ \\ \hline \hline
1   & 1            & 11  & 0.709998411 \\ \hline
2   & 0.840896415  & 12  & 0.708235422 \\ \hline
3   & 0.781934407  & 13  & 0.706789935 \\ \hline
4   & 0.757237797  & 14  & 0.705659505 \\ \hline
5   & 0.740759114  & 15  & 0.704734605 \\ \hline
6   & 0.731210449  & 16  & 0.704006945 \\ \hline
7   & 0.723927185  & 17  & 0.703419862 \\ \hline
8   & 0.719045188  & 18  & 0.702964717 \\ \hline
9   & 0.715146335  & 19  & 0.702610220 \\ \hline
10  & 0.712308458  & 20  & 0.702349024 \\ \hline
\end{tabular}
\end{center}
\caption[The first 20 Pad\'e approximants for the instanton problem]
{\label{tab}
The first 20 Pad\'e approximants for the solution to the instanton problem
(\ref{BL11}).}
\end{table}


\begin{table}[htbp]
\begin{center}
\begin{tabular}{|c|l||c|l|} \hline
$N$ & $b_0^{(N)}$  & $N$ & $b_0^{(N)}$ \\ \hline \hline
180   & 0.707530492  & 190  & 0.707471024 \\ \hline
181   & 0.707524250  & 191  & 0.707465419 \\ \hline
182   & 0.707518076  & 192  & 0.707459872 \\ \hline
183   & 0.707511970  & 193  & 0.707454384 \\ \hline
184   & 0.707505930  & 194  & 0.707448952 \\ \hline
185   & 0.707499955  & 195  & 0.707443575 \\ \hline
186   & 0.707494044  & 196  & 0.707438253 \\ \hline
187   & 0.707488197  & 197  & 0.707432986 \\ \hline
188   & 0.707482412  & 198  & 0.707427771 \\ \hline
189   & 0.707476687  & 199  & 0.707422609 \\ \hline
\end{tabular}
\end{center}
\caption[The last 20 variational strong-coupling coefficients for the
instanton
problem]
{\label{VPTapp} The last 20 variational strong-coupling coefficients
$b_0^{(N)}$ from Eq.~(\ref{BL18}).}
\end{table}

\begin{table}[htbp]
\begin{center}
\begin{tabular}{|l|l|l|} \hline
order & value for $b_0^{(N)}$ & convergence\\ \hline \hline
1   & 0.70640049   & decreasing   \\ \hline
2   & 0.70639983200   & increasing \\ \hline
3   & 0.706399832082   & increasing \\ \hline
4   & 0.7063998320858658 & increasing \\ \hline
5   & 0.706399832085884411 & increasing \\ \hline
6   & 0.70639983208588446498 & increasing \\ \hline
\end{tabular}
\end{center}
\caption[Six orders of Richardson extrapolation for the strong-coupling
coefficient $b_0^{(N)}(k_0)$ for the instanton problem]
{\label{VPTtab}
Six orders of Richardson extrapolations for the strong-coupling coefficient
$b_0^{(N)}(k_0)$ up to $N=200$ for the instanton problem. The last value is
only
$0.099\%$ away from the correct limiting value
$1/\sqrt{2}=0.7071067812...\,.$}
\end{table}

\begin{table}[htbp]
\begin{center}
\begin{tabular}{|l|l|l|} \hline
order & value for $A$ & convergence\\ \hline \hline
1   & -1.4998    & increasing \\ \hline
2   & -1.500017  & decreasing \\ \hline
3   & -1.5000011 & decreasing \\ \hline
4   & -1.49999874 & increasing \\ \hline
5   & -1.5000004  & decreasing \\ \hline
6   & -1.499999893 & increasing \\ \hline
\end{tabular}
\end{center}
\caption[Six orders of Richardson extrapolation for the exponent $A$ for the
large-order instanton weak-coupling coefficients]
{\label{richy2}
Six orders of Richardson extrapolations for the exponent $A$ of the
large-order
instanton weak-coupling coefficients, based on the first 200 weak-coupling
coefficients. The value $A=-3/2$ is quite plausible.}
\end{table}


\begin{table}[htbp]
\begin{center}
\begin{tabular}{|l|l|l|} \hline
order & value for $K$ & convergence\\ \hline \hline
1   & 2.46692    & decreasing \\ \hline
2   & 2.4668283  & increasing \\ \hline
3   & 2.46682911 & decreasing \\ \hline
4   & 2.466829065 & decreasing \\ \hline
5   & 2.4668290597 & increasing \\ \hline
6   & 2.4668290635 & decreasing \\ \hline
\end{tabular}
\end{center}
\caption[Six orders of Richardson extrapolation for the inverse radius of
convergence $K$ of the large-order instanton weak-coupling coefficients]
{\label{richy3}
Six orders of Richardson extrapolations for the inverse radius of
convergence
$K$ of the large-order instanton weak-coupling coefficients, based on the
first
200 weak-coupling coefficients under the assumption that $A=-3/2$.}
\end{table}

\begin{table}[htbp]
\begin{center}
\begin{tabular}{|l|l|l|} \hline
order & value for $B_1$ & convergence \\ \hline \hline
1   & 0.0170837  & increasing \\ \hline
2   & 0.0170864  & increasing \\ \hline
3   & 0.017087  & increasing \\ \hline
4   & 0.0170893 & increasing \\ \hline
5   & 0.0170908 & increasing \\ \hline
6   & 0.0170922 & increasing \\ \hline
\end{tabular}
\end{center}
\caption[Six orders of Richardson extrapolation for the overall factor $B_1$
of
the large-order instanton weak-coupling coefficients]
{\label{richy4}
Six orders of Richardson extrapolations for the overall factor $B_1$ of the
large-order instanton weak-coupling coefficients, based on the first 200
weak-coupling coefficients under the assumption that $K=2.4482906$ and $A=-
3/2$. The value of $B_1$ strongly depends on the numerical values for $A$
and
$K$. Changing $K$ in the sixth decimal place influences the third
significant
figure of $B_1$. Also, all the Richardson extrapolations are increasing so,
strictly speaking, we only have a lower boundary for $B_1$. Thus, the
accuracy
of $B_1$ may not be very good.}
\end{table}

\begin{table}[htbp]
\begin{center}
\begin{tabular}{|l|l|l|} \hline
order & value for $B_2$ & convergence \\ \hline \hline
1   & 0.119069  & increasing \\ \hline
2   & 0.119083  & increasing \\ \hline
3   & 0.119093  & increasing \\ \hline
4   & 0.119054095 & increasing \\ \hline
5   & 0.119054125 & increasing \\ \hline
6   & 0.119054146 & increasing \\ \hline
\end{tabular}
\end{center}
\caption[Six orders of Richardson extrapolation for the overall factor $B_2$
of
the large-order instanton weak-coupling coefficients]
{\label{richy5}
Six orders of Richardson extrapolations for the overall factor $B_2$ of the
large-order instanton weak-coupling coefficients based on the first 200
weak-coupling coefficients and the same assumptions as in the case of $B_1$
(see
Table \ref{richy4}). The value of $B_2$ depends strongly on $A$ and $K$.}
\end{table}

\begin{table}[htbp]
\begin{center}
\begin{tabular}{|c|c||c|c|} \hline
$j$ & $a_{1,j}$ & $j$ & $a_{1,j}$ \\[1mm] \hline \hline
\rule[-5pt]{0pt}{20pt}
1 & $-2$ & 11  & $\frac{30868632383}{5457375}$ \\[1mm] \hline
\rule[-5pt]{0pt}{20pt}
2 & $2$  & 12  & $\frac{6325029622}{637875}$ \\[1mm] \hline
\rule[-5pt]{0pt}{20pt}
3 & $\frac{8}{3}$  & 13  & $-\frac{487693745019181}{13408770375}$ \\[1mm]
\hline
\rule[-5pt]{0pt}{20pt}
4 & $-6$ & 14  & $-\frac{4774319527974167}{37819608750}$ \\[1mm] \hline
\rule[-5pt]{0pt}{20pt}
5 & $-\frac{184}{15}$ & 15  & $\frac{430321251088745734}{2212447111875}$
\\[1mm] \hline
\rule[-5pt]{0pt}{20pt}
6 & $\frac{136}{9}$ & 16  & $\frac{796235344548876790517}{603998061541875}$
\\[1mm] \hline
\rule[-5pt]{0pt}{20pt}
7 & $\frac{11062}{105}$ & 17  &
$-\frac{2249988054506764174584049}{6776858250499837500}$ \\ \hline
\rule[-5pt]{0pt}{20pt}
8 & $-\frac{8162}{225}$ & 18  &
$-\frac{178060537619150189817796}{14237097164915625}$ \\[1mm] \hline
\rule[-5pt]{0pt}{20pt}
9 & $-\frac{10557416}{14175}$  & 19  &
$-\frac{13224896152219729667498038639}{1301909768346024337500}$ \\[1mm]
\hline
\rule[-5pt]{0pt}{20pt}
10 & $-\frac{57628622}{99225}$ & 20  &
$\frac{121756993154067534451733120837029}{1153217968487557347375000}$
\\[1mm] \hline
\end{tabular}
\end{center}
\caption[The first 20 weak-coupling coefficients for the Blasius equation
(non-Borel)]
{\label{tabelle}
The first 20 weak-coupling coefficients for the Blasius recursion relation
(\ref{blasius4}) and (\ref{hmmm}).
Observe that the coefficients $a_{1,j}$ are not of Borel type
(they do not alternate in sign). A cosine function with a frequency
different
from $\pi$ governs the sign pattern (see Sec.~\ref{sec8}).}
\end{table}

\begin{table}[htbp]
\begin{center}
\begin{tabular}{|c|l||c|l|} \hline
$N$ & $S_N$         & $N$ & $S_N$ \\ \hline \hline
1   & 0.5           & 11  & 0.3574632121 \\ \hline
2   & 0.4204482076  & 12  & 0.3563326651 \\ \hline
3   & 0.3948201830  & 13  & 0.3553848048 \\ \hline
4   & 0.3819443732  & 14  & 0.3545795944 \\ \hline
5   & 0.3742062309  & 15  & 0.3538882842 \\ \hline
6   & 0.3690504811  & 16  & 0.3532891509 \\ \hline
7   & 0.3653779673  & 17  & 0.3527655813 \\ \hline
8   & 0.3626359060  & 18  & 0.3523046588 \\ \hline
9   & 0.3605155915  & 19  & 0.3518961929 \\ \hline
10  & 0.3588309707  & 20  & 0.3515320399 \\ \hline
\end{tabular}
\end{center}
\caption[The first 20 Pad\'e approximants for the Blasius equation]
{\label{tabelle2}
The first 20 Pad\'e approximants for the solution to the Blasius equation
(\ref{blasius1}). The sequence formed by the $S_N$ converges extremely
slowly.}
\end{table}

\begin{table}[htbp]
\begin{center}
\begin{tabular}{|l|l|l|} \hline
order & value of $y''(0)$ & convergence \\ \hline \hline
1   & 0.3445  & decreasing \\ \hline
2   & 0.3436  & decreasing \\ \hline
3   & 0.3430  & oscillating \\ \hline
\end{tabular}
\end{center}
\caption[Three orders of Richardson extrapolation for the Blasius equation]
{\label{richy}
Three orders of Richardson extrapolations for the Blasius equation
(\ref{blasius1}), based on the first 70 Pad\'e approximants $S_N$.}
\end{table}

\begin{table}[htbp]
\begin{center}
\begin{tabular}{|c|l||c|l|} \hline
$N$ & $b_0^{(N)}$  & $N$ & $b_0^{(N)}$ \\ \hline \hline
180   & 0.33696017793094  & 190  & 0.33695971119646 \\ \hline
181   & 0.33696012777085  & 191  & 0.33695966849139 \\ \hline
182   & 0.33696007843082  & 192  & 0.33695962644843 \\ \hline
183   & 0.33696002989308  & 193  & 0.33695958505396 \\ \hline
184   & 0.33695998214034  & 194  & 0.33695954429471 \\ \hline
185   & 0.33695993515575  & 195  & 0.33695950415774 \\ \hline
186   & 0.33695988892292  & 196  & 0.33695946463046 \\ \hline
187   & 0.33695984342591  & 197  & 0.33695942570058 \\ \hline
188   & 0.33695979864918  & 198  & 0.33695938735612 \\ \hline
189   & 0.33695975457760  & 199  & 0.33695934958540 \\ \hline
\end{tabular}
\end{center}
\caption[The last 20 variational strong-coupling coefficients for the
Blasius
equation]
{\label{bald} The last 20 variational
strong-coupling coefficients $b_0^{(N)}$
for the Blasius equation. The very last coefficient is
$b^{(200)}_0=0.33695931237713$, as opposed to the correct value
$y''(0)=0.33206$.}
\end{table}

\begin{table}[htbp]
\begin{center}
\begin{tabular}{|l|l|l|} \hline
order & value for $b_0^{(N)}$ & convergence\\ \hline \hline
1   & 0.3369518           & increasing \\ \hline
2   & 0.336955563         & increasing \\ \hline
3   & 0.336955600539      & increasing \\ \hline
4   & 0.3369556008803     & increasing \\ \hline
5   & 0.336955600883462   & increasing \\ \hline
6   & 0.33695560088349232 & increasing \\ \hline
\end{tabular}
\end{center}
\caption[Six orders of Richardson extrapolation for the strong-coupling
coefficient $b_0^{(N)}(k_0)$ for the Blasius equation]
{\label{bald2}
Six orders of Richardson extrapolations for the strong-coupling coefficient
$b_0^{(N)}(k_0)$ up to $N=200$ for the Blasius equation. The last value is
$1.5\%$ away from the correct limiting value $y''(0)=0.33206$.}
\end{table}

\begin{table}[htbp]
\begin{center}
\begin{tabular}{|l|l|} \hline
$a$ & $b$          \\ \hline \hline
1.3941    & 3.09   \\ \hline
1.3939    & 3.11   \\ \hline
7.67830   & 3.031  \\ \hline
7.67686   & 3.130  \\ \hline
\end{tabular}
\end{center}
\caption[Parameters for the large-order behavior of the Blasius
weak-coupling
coefficients]
{\label{tabelle3}
Examples of the parameters $a$ and $b$ that give the first 300 signs of the
Blasius weak-coupling coefficients correctly, assuming that the sign
structure
of the underlying large-order behavior is of the form $\cos (an+b)$. The
last
two values for $a$ can be obtained approximately by summing $2\pi$ to the
first
two values.}
\end{table}

\newpage
\typeout{==> Figures}
\section*{Figures}

\begin{thebibliography}{100}

\bibitem{Ben1} C.~M.~Bender, F.~Cooper, G.~S.~Guralnik, and D.~H.~Sharp,
Phys.~Rev.~D {\bf 19}, 1865 (1979).

\bibitem{Ben2} C.~M.~Bender, F.~Cooper, G.~S.~Guralnik, R.~Roskies, and
D.~H.~Sharp, Phys. Rev.~Lett.~{\bf 43}, 537 (1979).

\bibitem{Ben3}
C.~M.~Bender, F.~Cooper, G.~S.~Guralnik, H.~Rose, and D.~H.~Sharp,
Jour.~Stat.~Phys. {\bf 22}, 647 (1980).

\bibitem{Ben4} C.~M.~Bender, F.~Cooper, G.~S.~Guralnik, R.~Roskies, and
D.~H.~Sharp, in {\it Recent Developments in High-Energy Physics}, Ed.~by
B.~Kursunoglu, A.~Perlmutter, and L.~F.~Scott (Plenum, New York, 1980),
p.~211.

\bibitem{Ben5} C.~M.~Bender, F.~Cooper, R.~Kenway, and L.~M.~Simmons, Jr.,
Phys.~Rev.~D {\bf 24}, 2693 (1981).

\bibitem{Ben6} C.~M.~Bender and R.~Z.~Roskies,
Phys.~Rev.~D {\bf 25}, 427 (1982).

\bibitem{Ben7} C.~M.~Bender, F.~Cooper, R.~Kenway, and L.~M.~Simons, Jr.,
Phys.~Lett {\bf 109 B}, 63 (1982).

\bibitem{Ben8} C.~M.~Bender, L.~R.~Mead, and L.~M.~Simmons, Jr.,
Phys.~Rev.~D {\bf 28}, 936 (1983).

\bibitem{Ben9} C.~M.~Bender, F.~Cooper, and A.~Das,
Phys.~Rev.~Lett.~{\bf 50}, 397 (1983).

\bibitem{Bender1} C.~M.~Bender, F.~Cooper, G.~S.~Guralnik, E.~Mjolsness,
H.~A.~Rose, and D.~H.~Sharp, Adv.~Appl.~Mathematics {\bf 1}, 22 (1980).

\bibitem{Bender2} C.~M.~Bender, Los Alamos Science {\bf 2}, 76 (1981).

\bibitem{Bender3} C.~M.~Bender and A.~Tovbis,
J.~Math.~Phys.~{\bf 38}, 3700 (1997).

\bibitem{BO} C.~M.~Bender and S.~A.~Orszag, {\it Advanced Mathematical
Methods
for Scientists and Engineers} (McGraw-Hill, New York, 1978).

\bibitem{internet2} The first 200 weak-coupling coefficients for the
instanton
equation can be found at
{\tt http://www.physik.fu-berlin.de/\~{}weissbach/inst.html $\,$.}

\bibitem{trick} H.~Kleinert, {\it Path integrals in quantum mechanics,
statistics, and polymere physics}, second ed.
(World Scientific, Singapore, 1995).

\bibitem{Frohlinde} H.~Kleinert and V.~Schulte-Frohlinde, {\it Critical
Properties of $\phi^4$ Theories} (World Scientific, Singapore, 2001).

\bibitem{Stevenson}
P.~M.~Stevenson, Phys.~Rev.~D {\bf 23}, 2916 (1981)

\bibitem{blasius} H.~Blasius, Z.~Math.~Phys.~{\bf 56}, 1 (1908).

\bibitem{acta} Z.~Belhachmi, B.~Brighi, and K.~Taous, Acta
Math.~Univ.~Comenianae, Vol.~{\bf LXIX}, 199 (2000).

\bibitem{internet3}
The first 300 weak-coupling coefficients for the Blasius equation can be
found
at {\tt http://www.physik.fu-berlin.de/\~{}weissbach/blas.html $\,$.}

\end{thebibliography}
\end{document}